\documentclass[runningheads,a4paper]{llncs}

\usepackage[english]{babel}
\usepackage[utf8]{inputenc}   %%  8 bits using UTF-8

\usepackage{graphicx}
\usepackage{epstopdf}
\graphicspath{{figures/}} % declare the path(s) where your graphic files are

\usepackage{tabularx}
\usepackage[table]{xcolor}
\usepackage{multicol}
\usepackage{multirow}
\usepackage{booktabs}
\usepackage{framed}

\usepackage{rotating}

\definecolor{gray}{rgb}{0.5,0.5,0.5}
\definecolor{mauve}{rgb}{0.58,0,0.82}
\definecolor{rltred}{rgb}{0.5,0,0}
\definecolor{rltgreen}{rgb}{0,0.5,0}
\definecolor{rltblue}{rgb}{0,0,0.5}

\usepackage{boxedminipage}
\newenvironment{result}%
{\smallskip
\noindent
\let\emph=\textbf
\begin{boxedminipage}{\columnwidth}\begin{center}\em}%
{\end{center}\end{boxedminipage}%
}

\usepackage{theorem}
\usepackage{amsmath}
\usepackage{amssymb}
\usepackage{bm}
\usepackage{xfrac}
\usepackage{xspace}

\usepackage{paralist}
\usepackage{verbatim}
\usepackage{listings}

\usepackage{algpseudocode}
\usepackage{algorithm}

\usepackage{setspace}

\usepackage{subcaption}
\usepackage{graphicx}

\usepackage{float}

\usepackage{placeins} % for FloatBarrier

\usepackage{url}
\usepackage{hyperref}
\usepackage[hang,flushmargin]{footmisc}

\newif\ifdraft\drafttrue
\drafttrue

\ifdraft
	\newcommand\TODO[1]{\textcolor{rltred}{\textbf{\colorbox{yellow}{\small TODO:}} \emph{#1}}\xspace}
	\newcommand\DONE[1]{\textcolor{rltgreen}{\textbf{\colorbox{green}{\small Done:}} \emph{#1}}\xspace}
\else
  \newcommand\TODO[1]{}
  \newcommand\DONE[1]{}
\fi

\usepackage{cite}
\usepackage{balance}

\usepackage{lipsum} % just to automatically generate some text
\newcommand{\keywords}[1]{\par\addvspace\baselineskip
\noindent\keywordname\enspace\ignorespaces#1}

\DeclareMathSymbol{,}{\mathpunct}{letters}{"3B}
\DeclareMathSymbol{,}{\mathord}{letters}{"3B}
\DeclareMathSymbol{\decimal}{\mathord}{letters}{"3A}
\begin{document}

\mainmatter % start of an individual contribution

\title{Many Independent Objective (MIO) Algorithm for Test Suite Generation}
%\title{Massive Independent Objective (MIO) Algorithm for Test Suite Generation}

\author{
Andrea Arcuri%\inst{1}
}
\authorrunning{}

\institute{
Westerdals Oslo ACT, Faculty of Technology, Oslo, Norway, \\ and University of Luxembourg, Luxembourg\\
}

\toctitle{}
\tocauthor{}
\maketitle

% ----------------------------------------------------------- Abstract

\begin{abstract}
Automatically generating test suites is intrinsically a multi-objective problem, as any of the testing targets (e.g, statements to execute or mutants to kill) is an objective on its own. 
Test suite generation has peculiarities that are quite different from other more regular optimisation problems.
For example, given an existing test suite, one can add more tests to cover the remaining objectives.
One would like the smallest number of small tests to cover as many objectives as possible, but that is a secondary goal compared to covering those targets in the first place.
Furthermore, the amount of objectives in software testing can quickly become unmanageable, in the order of (tens/hundreds of) thousands, especially for system testing of industrial size systems.
Traditional multi-objective optimisation  algorithms can already start to struggle with just four or five objectives to optimize.
To overcome these issues,  different techniques have been proposed, like for example the Whole Test Suite (WTS) approach and the Many-Objective Sorting Algorithm (MOSA).
However, those techniques might not scale well to very large numbers of objectives and  limited search budgets (a typical case in system testing).
In this paper, we propose a novel algorithm, called Many Independent Objective (MIO) algorithm. 
This algorithm is designed and tailored based on the specific properties of test suite generation.
An empirical study, on a set of artificial and actual software, shows that the MIO algorithm can achieve higher coverage compared to WTS and MOSA, as it can better exploit the peculiarities of test suite generation.
\end{abstract}

%%%%%%%%%%%%%%%%%%%%%%%%%%%%%%%%%%%%%%%%%%%%%%%%%%%%%%%%%%%%%%%%%%%%%%%%%
\section{Introduction}
%\vspace{-0.5em}

Test case generation can be modelled as an optimisation problem, and so different kinds of search algorithms can be used to address it~\cite{harman2012search}.
There can be different objectives to optimise, like for example branch coverage or the detection of mutants in the system under test (SUT). 
When aiming at maximising these metrics, often the sought solutions are not single test cases, as a single test cannot cover all the objectives in the SUT.
Often, the final solutions are sets of test cases, usually referred as \emph{test suites}.

There are many different kinds of search algorithms that can be used for generating test suites.
The most famous is perhaps the Genetic Algorithms (GA), which is often the first choice when addressing a new software engineering problem for the first time.
But it can well happen that on specific problems other search algorithms could be better.
Therefore, when investigating a new problem, it is not uncommon to evaluate and compare different algorithms.
On average, no search algorithm can be best on all possible problems~\cite{WoM97}.
It is not uncommon that, even on non-trivial tasks, simpler algorithms like (1+1)~Evolutionary Algorithm (EA) or Hill Climbing (HC) can give better results than GA (e.g., as in~\cite{ali2013generating}).

A major factor affecting the performance of a search algorithm is the so called \emph{search budget}, i.e., for how long the search can be run, usually the longer the better.
But the search budget is also strongly related to the tradeoff between the \emph{exploitation} and \emph{exploration} of the search landscape.
If the budget is low, then a population-based algorithm like GA (which puts more emphasis on the exploration) is likely to perform worse than a single, more focused individual-based algorithm like HC or (1+1)~EA.
On the other hand, if the search budget is large enough, the exploration made by the GA can help it to escape from the so called local optima in which HC and (1+1)~EA can easily get stucked in.

To obtain even better results, then one has to design specialised search algorithms that try to exploit the specific properties of the addressed problem domain.
In the case of test suite generation, there are at least the following peculiarities:
\begin{itemize}
\item testing targets can be sought \emph{independently}. Given an existing test suite, to cover the remaining testing targets (e.g., lines and branches), you can create and add new tests without the need to modify the existing ones in the suite. 
At the end of the search, one wants a minimised test suite, but that is a secondary objective compared to code coverage.
\item testing targets can be strongly related (e.g., two nested branches), as well as being completely independent (e.g., code in two different top-level functions with no shared state). 
\item some testing targets can be \emph{infeasible}, i.e., impossible to cover. There can be different reasons for it, e.g., dead code, defensive programming or the testing tool not handling all kinds of SUT inputs (e.g., files or network connections). Detecting whether a target is feasible or not is an undecidable problem.
\item for non-trivial software, there can be a very large number of objectives. This is specially true not only for system-level testing, but also for unit testing when mutation score is one of the coverage criteria~\cite{mutation2015emse}.
Traditional multi-objective algorithms are ill suited to tackle large numbers of objectives~\cite{li2015many}. 
\end{itemize}

In this paper, we propose a novel search algorithm that exploits such characteristics and, as such, it is specialised for test suite generation (and any problem sharing those properties).
We call it the Many Independent Objective (MIO) algorithm.
We carried out an empirical study to compare the MIO algorithm with the current state-of-the-art, namely the Whole Test Suite~\cite{GoA_TSE12} approach and the Many-Objective Sorting Algorithm~\cite{dynamosa2017}.
On a set of artificial software with different characteristics (clear gradients, plateaus, deceptive local optima and infeasible targets), in most cases MIO achieves higher coverage.
This was also confirmed with unit test experiments on three numerical functions.

%%%%%%%%%%%%%%%%%%%%%%%%%%%%%%%%%%%%%%%%%%%%%%%%%%%%%%%%%%%%%%%%%%%%%%%%%
\section{Background}
%\vspace{-0.5em}

%%------------------------------------------------------------------------
%\subsection{Search-Based Software Testing}
%

%------------------------------------------------------------------------
\subsection{Whole Test Suite (WTS)}

The Whole Test Suite~\cite{GoA_TSE12} approach was introduced as an algorithm to generate whole test suites. 
Before that, typical test case generators were targeting only single objectives, like specific lines or branches, using heuristics like the \emph{branch distance} and the \emph{approach level} (as for example done in~\cite{HaM10}).

In the WTS approach, a GA is used, where an individual in the GA population is a set of test cases. 
Mutation and crossover operators can modify both the set composition (i.e., remove or add new tests) and its content (e.g., modify the tests).
As fitness function, the sum of all branch distances in the SUT is used.
At the end of the search, the best solution in the population is given as output test suite.
To avoid losing good tests during the search, the WTS can also be extended to use an \emph{archive} of best tests seen so far~\cite{rojas2016detailed}.

%------------------------------------------------------------------------
\subsection{Many-Objective Sorting Algorithm (MOSA)}

The Many-Objective Sorting Algorithm (MOSA)~\cite{dynamosa2017} was introduced to overcome some of the limitations of WTS.
In MOSA, each testing target (e.g., lines) is an objective to optimize.
MOSA is an extension of NSGA-II~\cite{deb2002fast}, a very popular multi-objective algorithm.
In MOSA, the population is composed of tests, not test suites. 
When a new target is covered, the test covering it gets stored in an archive, and such target is not used any more in the fitness function.
A final output test suite is composed by the best tests found during the search and that are stored in the archive.

In NSGA-II, selection is based on ranks (from 1 on, where 1 is the best): an individual that subsumes many other individuals gets a better rank, and so it is more likely to be selected for reproduction.
One of the main differences of MOSA compared to NSGA-II is the use of the \emph{preference sorting criterion}: to avoid losing the best individuals for a given testing target, for each uncovered testing target the best individual gets the best rank (0 in MOSA), regardless of its subsuming relations with the other tests.

%%%%%%%%%%%%%%%%%%%%%%%%%%%%%%%%%%%%%%%%%%%%%%%%%%%%%%%%%%%%%%%%%%%%%%%%%
\section{The MIO Algorithm}
%\vspace{-0.5em}

%------------------------------------------------------------------------
\subsection{Core Algorithm}
\label{sec:core}

Both WTS and MOSA have been shown to provide good results, at least for unit test generation~\cite{GoA_TSE12,rojas2016detailed,dynamosa2017}.
However, both algorithms have intrinsic limitations, like for example:
\begin{itemize}
\item	population-based algorithms like WTS and MOSA do put more emphasis on the  exploration of the search landscape, which is not ideal in constrained situations of limited search budgets, like for example in system-level testing where each test case execution can be computationally expensive.
Letting the user to tune the population size parameter is not a viable option, unless it is done automatically (but even then, it has side effects, as we will see in the empirical study).
\item although once a target is covered it is not used any more for the fitness function,  the individuals optimised for it would still be in the population.
They will die out eventually after a few generations, but, until then, their presence in the population can hamper the search if those covered targets are unrelated to the remaining non-covered targets.  
\item in the presence of infeasible targets, some tests can get good fitness score (e.g., a close to 0 branch distance) although they will never cover those infeasible targets. Those not useful tests might end up taking over a large part of the population.
\item there can be a very large number of objectives to cover, even in the order of hundreds of thousands (e.g., in the system-level testing of industrial systems). 
A fixed size population would simple not work well: if too small, then there would not be enough diverse genetic material in the first generation;
if too large, not only convergence would be drastically slowed down, but also the computational cost could sky-rock (e.g., NSGA-II has a quadratic complexity based on the population size).
\end{itemize}

To avoid these limitations, we have designed a novel evolutionary algorithm that we call the Many Independent Objective (MIO) algorithm.
In a nutshell, MIO combines the simplicity and effectiveness of (1+1) EA with a dynamic population, dynamic exploration/exploitation tradeoff and feedback-directed target selection.

The MIO algorithm maintains an archive of tests.
In the archive, \emph{for each} testing target we keep a different population of tests of size up to $n$ (e.g, $n=10$).
Therefore, given $z$ objectives/targets, there can be up to $n \times z$ tests in the archive at the same time.   

At the beginning of the search, the archive will be empty, and so a new test will be randomly generated.
From the second step on, MIO will decide to either sample a new test at random (probability $P_r$), or will choose (details later) one existing test in the archive (probability $1-P_r$), copy it, and mutate it.
Every time a new test is sampled/mutated, its fitness is calculated, and it will be saved in the archive if needed (details later).
At this point, we need to define how tests are saved in the archive, and how MIO samples from the archive.

When a test is evaluated, a copy of it might be saved in 0 or more of the $z$ populations in the archive, based on its fitness value.
For each target, there will be a heuristics score $h$ in $[0, 1]$, where 1 means that the target is covered, whereas 0 is the worst possible heuristics value.
For example, if the heuristics is the branch distance $d$, this can be mapped into $[0, 1]$ by using $h = 1/(1+d)$ (where $h=0$ if a branch was never reached and so the branch distance $d$ was not calculated).

For each target $k$, a test is saved in population $T_k$, with $|T_k| \le n$, if either:
\begin{itemize}
\item	if $h_k=0$, the test is not added regardless of the following conditions.
\item	if the target is covered, i.e. $h_k=1$, the test is added and that population is shrunk to one single individual, and it will never expand again (i.e., it will be always $|T_k|=1$). A new test can \emph{replace} the one in $T_k$ only if it is \emph{shorter} (which will depend on the problem domain, e.g. size measured in sequence of function calls in unit testing) or, if it is of the same size, then replace the current test only if the new test has better coverage on the other targets (i.e., sum of all the heuristics values on all targets).
\item	if the population is not full (i.e., $|T_k| < n$), then the test is added. Otherwise, if full (i.e., $|T_k| = n$), the test might replace the worst in the population, but only if not worse than it (but not necessarily better). This means no worse heuristic value or, if the same, no larger size.
\end{itemize}

The idea is that, for each target, we keep a population of candidate tests for it, for which we have at least some heuristics value. 
But once a target $k$ is covered, we just need to store the best test, and discard the rest.
Note: if a discarded test in $T_k$ was good for another target $j$, then it would be still stored in $T_j$ anyway, so it is not lost.

When MIO needs to sample one test from the archive instead of generating one at random, it will do the following:
\begin{itemize}
\item choose one target $k$ at random where $|T_k|>0$ and $k$ is not covered (i.e., no test has $h_k=1$). If all non-empty populations are for covered targets, then just choose $k$ randomly among them.
\item choose one test randomly from $T_k$.
\end{itemize}

By using this approach, we aim at sampling tests that have non-zero heuristics (and so guidance) for targets that are not covered yet.

%------------------------------------------------------------------------
\subsection{Exploration/Exploitation Control}
\label{sec:control}

In the MIO algorithm, the two main parameters for handling the tradeoff between exploration and exploitation of the search landscape are the probability $P_r$ of sampling at random and the population size $n$ per target.
Exploration is good at the beginning of the search, but, at the end, a more focused exploitation can bring better results.
Like in Simulated Annealing, we use an approach in which we gradually reduce the amount of exploration during the search.

We define with $F$ the percentage of time after which a focused search should start.
This means that, for some parameters like $P_r$ and $n$, we define two values: one for the start of the search (e.g., $P_r=0.5$ and $n=10$), and one for when the focused phase begins (i.e., $P_r=0$ and $n=1$).
These values will linearly increase/decrease based on the passing of time.
For example, if $F=0.5$ (i.e., the focused search starts after 50\% of the search budget is used), then after 30\% of the search, the value $P_r$ would decrease from $0.5$ to $0.2$.

Note, when during the search decreasing $n$ leads to some cases with $|T|>n$, then those populations are shrunk by removing the worst individuals in it. 
Once the focused search begins (i.e., $P_r=0$ and $n=1$), then MIO starts to resemble a parallel (1+1)~EA.

When dealing with many objectives, even if there is a clear gradient to cover them in the fitness landscape, there might be simply not enough time left to cover all of them.
In software testing, the final user is only interested in tests that do cover targets, and not in tests that are heuristically close to cover them (e.g., close to solve complex constraints in some branch predicates, but not there yet).
Therefore, between a test suite $A$ that is close to but does not cover 100 targets, and another one $B$ which does cover 1 target and is very far from covering the remaining 99, the final user would likely prefer $B$ over $A$.  

To take this insight into account, MIO tries to focus on just few targets at a time, instead of spreading its resources thin among all the left uncovered targets.
For example, in MIO there is an extra parameter $m$ which controls how many mutations and fitness evaluations should be done on the same individual before sampling a new one.
Like $P_r$ and $n$, $m$ varies over time, like starting from $1$ and then increasing up to $10$ when the focused search begins.

%------------------------------------------------------------------------
\subsection{Feedback-Directed Sampling}
\label{sec:fds}

When dealing with many objectives and limited resources, it might not be possible to cover all of them.
As discussed in Section~\ref{sec:control}, the final user is only interested in the actually covered targets, and not on how close we are to cover them. 
Therefore, it makes sense to try to focus on targets that we have higher chances to cover.
This is helpful also when dealing with infeasible targets for which any heuristics will just plateau at a certain point.

To handle these cases, we use a simple but yet very effective technique that we call Feedback-Directed Sampling (FDS).
The sampling algorithm from the archive discussed in Section~\ref{sec:core} is modified as follow.
Instead of choosing the target $k$ randomly among the non-covered/non-empty ones, each of these targets will have a counter $c_k$.
Every time a test is sampled from a population $T_k$, then $c_k$ is increased by 1.
Every time a new \emph{better} individual is added to $T_k$ (or replace one of its existing tests), then the counter $c_k$ is reset to 0. 
When we sample from $k$ from non-covered/non-empty ones, then, instead of choosing $k$ at random, we choose the $k$ with the lowest $c_k$.

As long as we get improvements for a target $k$, the higher chances will be that we sample from $T_k$, as $c_k$ gets reset more often.
On the other hand, for infeasible targets, their $c$ will never be reset once they reach their plateau, and so they will be sampled less often.
Similarly, more complex targets will be sampled less often, and so the search concentrates on the easier targets that are not covered yet.
However, this is not an issue because, once an easy target $k$ is covered, we do not sample from $T_k$ any more (recall Section~\ref{sec:core}), unless also \emph{all} the other targets are either covered or with empty $T$.

%%%%%%%%%%%%%%%%%%%%%%%%%%%%%%%%%%%%%%%%%%%%%%%%%%%%%%%%%%%%%%%%%%%%%%%%%
\section{Empirical Study}
%\vspace{-0.5em}

To evaluate the performance of the MIO algorithm, we compared it with random search, MOSA and WTS.
We used two different case studies: 
(1) a set of artificial problems with varying, specific characteristics;
(2) three numerical functions.

In this paper, we aim at answering the following research questions:
\begin{description}
\item[{\bf RQ1}:] On which kinds of problem does MIO perform better than Random, MOSA and WTS?
\item[{\bf RQ2}:] What is the impact of tuning parameters for exploration vs.~exploitation of the search landscape in MIO and MOSA?
\item[{\bf RQ3}:] How do the analysed algorithms fare on actual software?
\end{description}

%-----------------------------------------------------------------------
\subsection{Artificial Software}

In this paper, we designed four different kinds of artificial problems.
In all of them, there are $z$ targets, and the search algorithm can be run for up to $b$ fitness evaluations.
A test is defined by two components: an $id$ (e.g., think about it like the name of a method to call in unit testing) and a numeric integer value $x \in [0, r]$ (e.g., think about it like the input to a method call).
Each target $k$ is independent, and can be covered only by a test with $id=k$.
The artificial problems will differ based on their fitness landscape.
Given $g \in [0, r]$ the single global optimum chosen at random, 
and given the normalising function $\rho(d) = 1 / (1 + d)$ for distances,
then we have four different cases for each target:
\begin{description}
\item[Gradient:]  $h_k = \rho(|x-g|)$. This represents the simplest case where the search algorithm has a direct gradient from $x$ toward the global optimum $g$.
\item[Plateau:] $h_k = \rho(g-x)$ if $g \ge x$, else $h_k = \rho(0.1 \times r)$.
In this case, we have one side of the search landscape (before the value of the global optimum $g$) with a clear gradient. However, the other side is a plateau with a relatively good fitness value (note that $0 \le |g-x| \le r$).
\item[Deceptive:] $h_k = \rho(g-x)$ if $g \ge x$, else $h_k = \rho(1 + r - x)$.
This is similar to the \emph{Plateau} case, where one side of the search landscape has a clear gradient toward $g$. However, the other side has a deceptive gradient toward leaving $g$ and reach the maximum value $r$.
\item[Infeasible:] like \emph{Gradient}, but with a certain number of the $z$ targets having a constant $h_k = \rho(1)$ and no global optimum.
\end{description}

We implemented the four search algorithms in which, when a test is sampled, its $id$ and $x$ values are chosen at random within the given valid ranges.
Mutations on $x$ is done by adding/subtracting $2^i$, where $i$ is chosen randomly in $[0, 10]$.
We consider mutating $id$ as a \emph{disruptive} operation, and, as such, we only mutate it with low probability $0.01$.
Mutating $id$ means changing both $id$ and $x$ at random (think about mutating a function call with string inputs into another one that requires integers, where the strings $x$ would have no meaning as integers).
All the analysed search algorithms use the same random sampling, mutation operation and archive to store the best tests found so far.

For the MIO algorithm, we used $F=0.5$, $P_r=0.5$, $n=10$ and max mutations 10.
For MOSA, we used the same settings as in~\cite{dynamosa2017}, i.e. population size 50
and tournament selection size 10.
WTS uses the same population size as MOSA, with up to 50 test cases in the same test suite (i.e., one individual).
A randomly sampled test suite in WTS will have size randomly chosen between 1 and 50.
WTS also has mutation operators to add a new test (probability 1/3) in a test suite, remove one test at random (probability 1/3), or modify one (probability 1/3) like in MIO and MOSA.
WTS also uses a crossover operator with probability 70\% to combine test suites.

\begin{figure}[!t]
  \centering
  \includegraphics[height=.40\textheight]{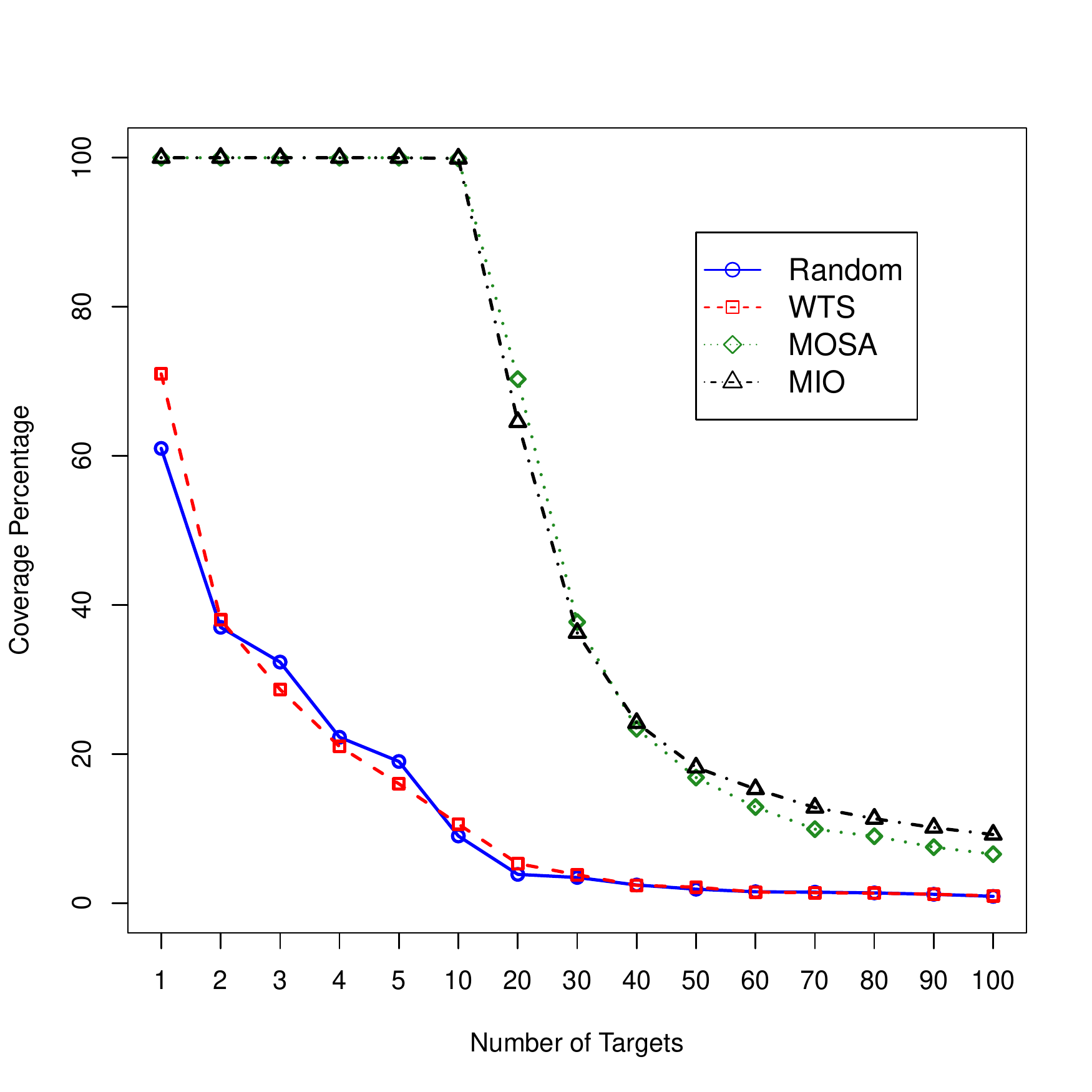}
  \caption{\label{fig:gradient}
  	Coverage results on the \emph{Gradient} problem type, with varying number of targets $z$.
  }
\end{figure}

\begin{figure}[!t]
  \centering
  \includegraphics[height=.40\textheight]{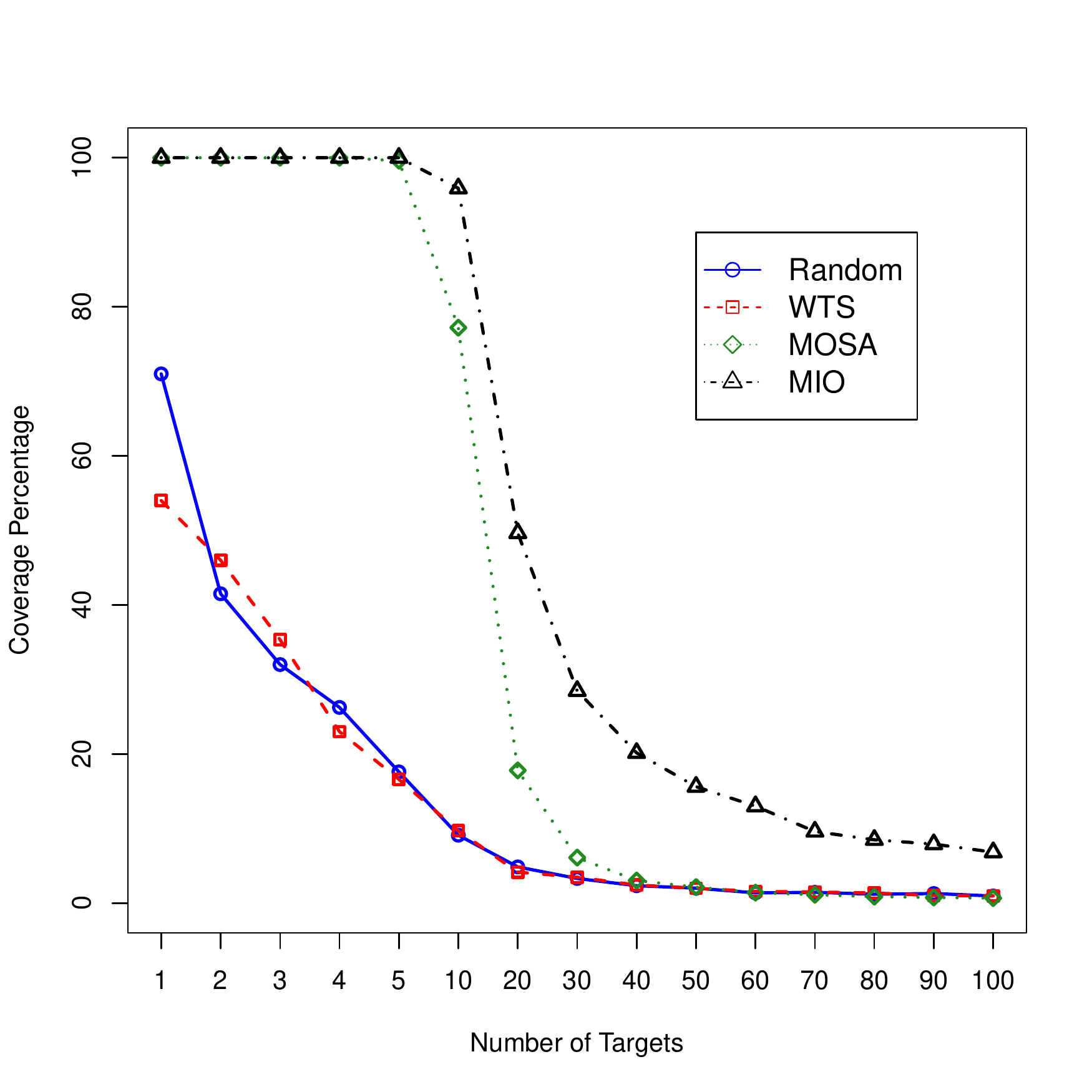}
  \caption{\label{fig:plateau}
  	Coverage results on the \emph{Plateau} problem type, with varying number of targets $z$.
  }
\end{figure}

\begin{figure}[!t]
  \centering
  \includegraphics[height=.40\textheight]{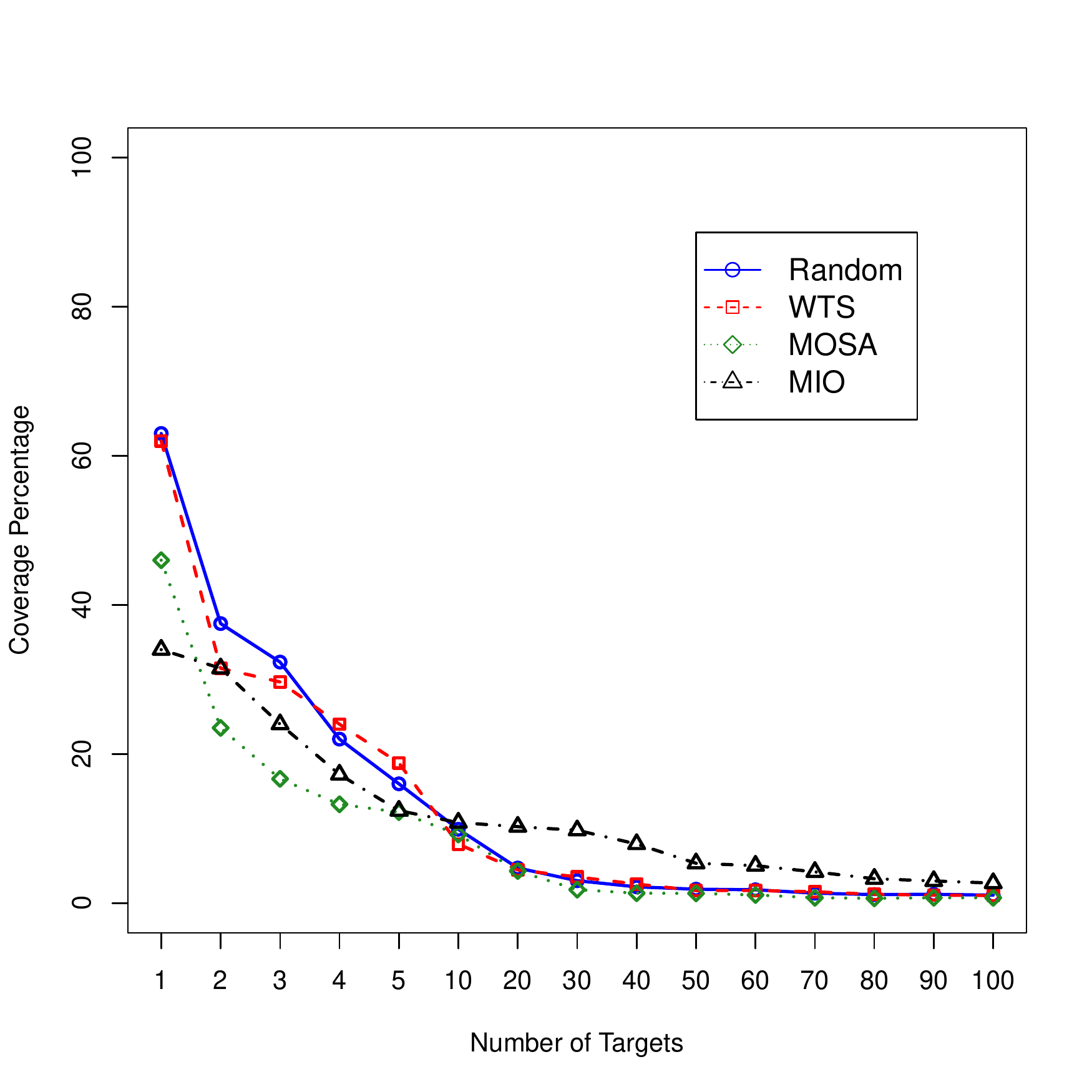}
  \caption{\label{fig:deceptive}
  	Coverage results on the \emph{Deceptive} problem type, with varying number of targets $z$.
  }
\end{figure}

\begin{figure}[!t]
  \centering
  \includegraphics[height=.40\textheight]{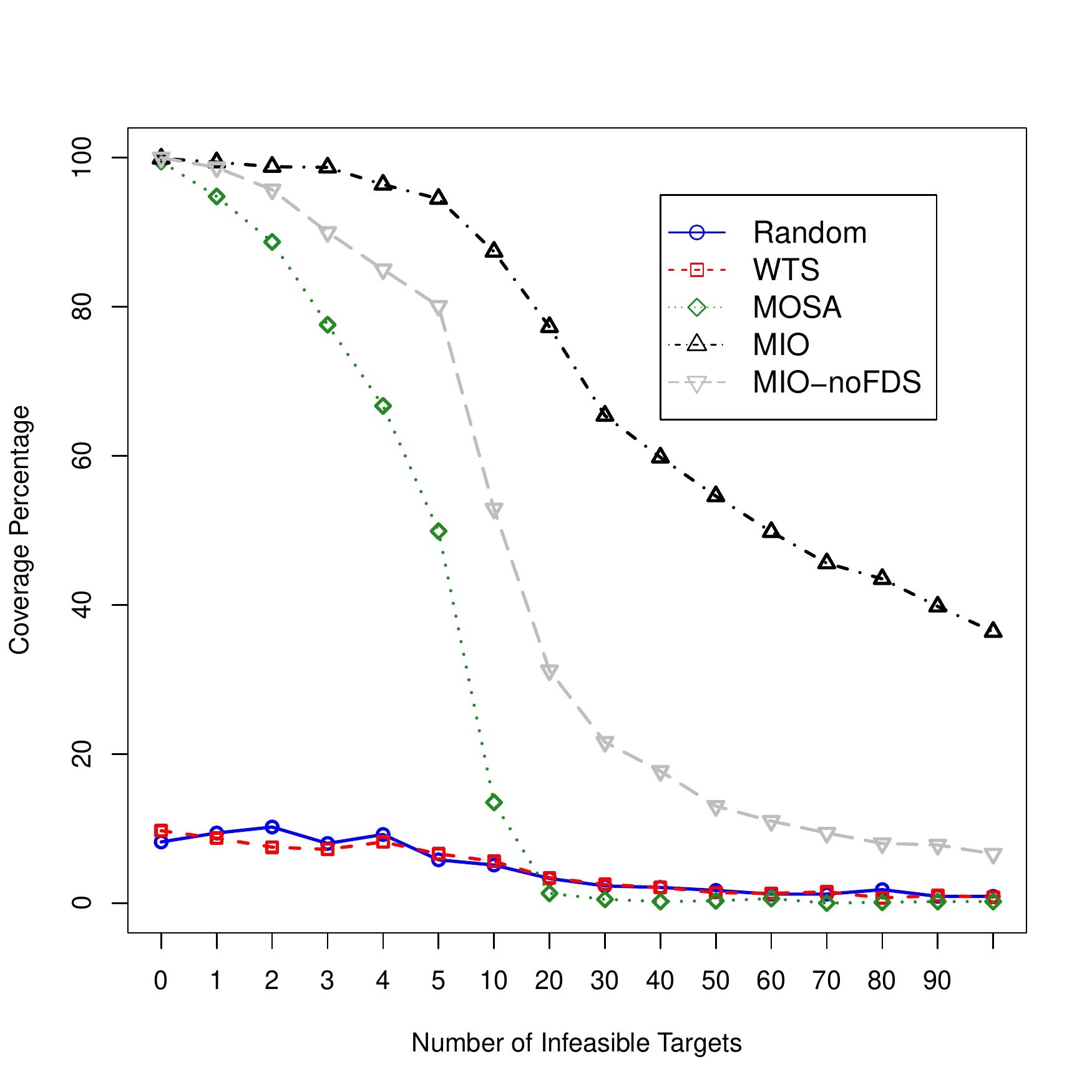}
  \caption{\label{fig:infeasible}
  	Coverage results on the \emph{Infeasible} problem type, with varying number of infeasible targets on top of 10 \emph{Gradient} ones.
  }
\end{figure}

For each problem type but \emph{Infeasible}, we created problems having a variable number of $z$ targets, in particular $z \in \{1, 2, 3, 4, 5, 10, 20, 30, 40, 50, 60, 70, 80, 90, 100\}$, i.e., 15 different values in total, ranging from 1 to 100.
We used $r=1000$.
We ran each of the four search algorithms $100$ times with budget $b=1000$.
As the optima $g$ are randomised, we make sure that the search algorithms run on the same problem instances.
In the case of the \emph{Infeasible} type, we used 10 \emph{Gradient} targets, on which we added a different number of infeasible targets in $\{0, 1, 2, 3, 4, 5, 10, 20, 30, 40, 50, 60, 70, 80, 90, 100\}$, i.e., 16 values in total, with $z$ ranging from $(10+0)=10$ to $(10+100)=110$.
Figure~\ref{fig:gradient} shows the results for the \emph{Gradient} type, 
Figure~\ref{fig:plateau} for \emph{Plateau},
Figure~\ref{fig:deceptive} for \emph{Deceptive},
and Figure~\ref{fig:infeasible} for \emph{Infeasible}.

The \emph{Gradient} case (Figure~\ref{fig:gradient})  is the simplest, where the four search algorithms obtain their highest coverage.
MIO and MOSA have very similar performance, which is higher than the one of Random and WTS.
However, on the more difficult case of \emph{Plateau} (Figure~\ref{fig:plateau}), MIO starts to have a clear advantage over MOSA.
For example, from $z=30$ on, MOSA becomes equivalent to Random and WTS, covering nearly no target. 
However, in that particular case, MIO can still achieve around 20\% coverage (i.e., 6 targets).
Even for large numbers of targets (i.e., 100 when taking into account that the search budget $b$ is only 1000), still MIO can cover some targets, whereas the other algorithms do not.

The \emph{Deceptive} case (Figure~\ref{fig:deceptive}) is of particular interest: for low numbers of $z$ targets (i.e., up to 10), both 
MIO and MOSA perform worse than Random. 
From 10 targets on, MOSA is equivalent to Random and WTS, whereas MIO has better results.
This can be explained by taking into account two contrasting factors:
(1) the more emphasis of MIO and MOSA on exploitation compared to the exploration of the search landscape is not beneficial in deceptive landscape areas, whereas a random search would not be affected by it;
(2) MIO does better handle large numbers of targets (Figure~\ref{fig:gradient}), even when there is no gradient (Figure~\ref{fig:plateau}). 
The value $z=10$ seems to be the turning point where (2) starts to have more weight than (1).

The \emph{Infeasible} case (Figure~\ref{fig:infeasible}) is where MIO obtains  the best results compared to the other algorithms.
For this case, we also ran a further version of MIO in which we deactivated FDS (recall Section~\ref{sec:fds}), as we wanted to study its impact in the presence of infeasible targets.
From 20 infeasible targets on, MOSA, Random and WTS become equivalent, covering nearly no target.
However, MIO can still cover nearly 80\% of the 10 feasible testing targets.
For very large numbers of infeasible targets like 100, still MIO can cover nearly 40\% of the feasible ones.  
This much better performance is mainly due to the use of FDS (see the gap in Figure~\ref{fig:infeasible} between MIO and MIO-noFDS).
However, even without FDS, MIO still does achieve better results compared to the other algorithms.

\begin{result}
{\bf RQ1}: on all the considered problems, MIO is the algorithm that scaled best. 
Coverage improvements were even up to 80\% in some cases.
\end{result}

When using a search algorithm, some parameters need to be set, like the population size or crossover probability in a GA.
Usually, common settings in the literature can already achieve good results on average~\cite{arcuri2013parameter}.
Finding tuned settings that work better on average on a large number of different artefacts is not trivial.
Ideally, a user should just choose for how long a search algorithm should run, and not do long tuning phases by himself on his problem artefacts.
Parameter tuning can also play a role in algorithm comparisons: what if a compared algorithm performed worse just because one of its chosen settings was sub-optimal?

Arguably, among the most important parameters for a search algorithm are the ones that most impact the tradeoff between the exploration and the exploitation of the search landscape.
In the case of MIO, this is clearly controlled by the $F$ parameter (low values put more emphasis on exploitation, whereas for high values a large number of tests are simply sampled at random).
In the case of population-based algorithms, the population size can be considered as a parameter to control such tradeoff.
Small populations would reward exploitation, whereas large populations would reward exploration.

\begin{figure}
  \centering
  \begin{subfigure}{\linewidth}
  	\includegraphics[width=.45\linewidth]{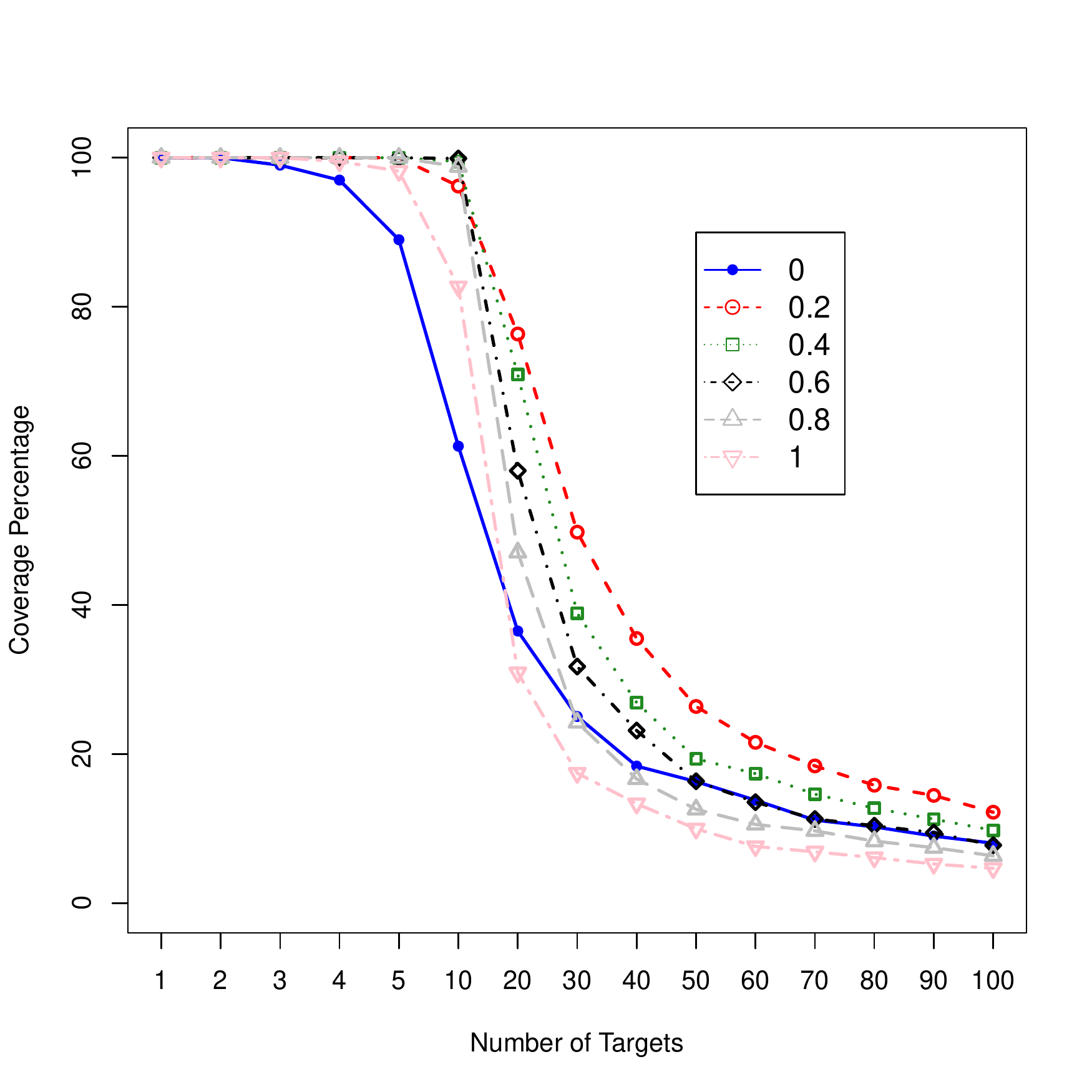}\hfill
  	\includegraphics[width=.45\linewidth]{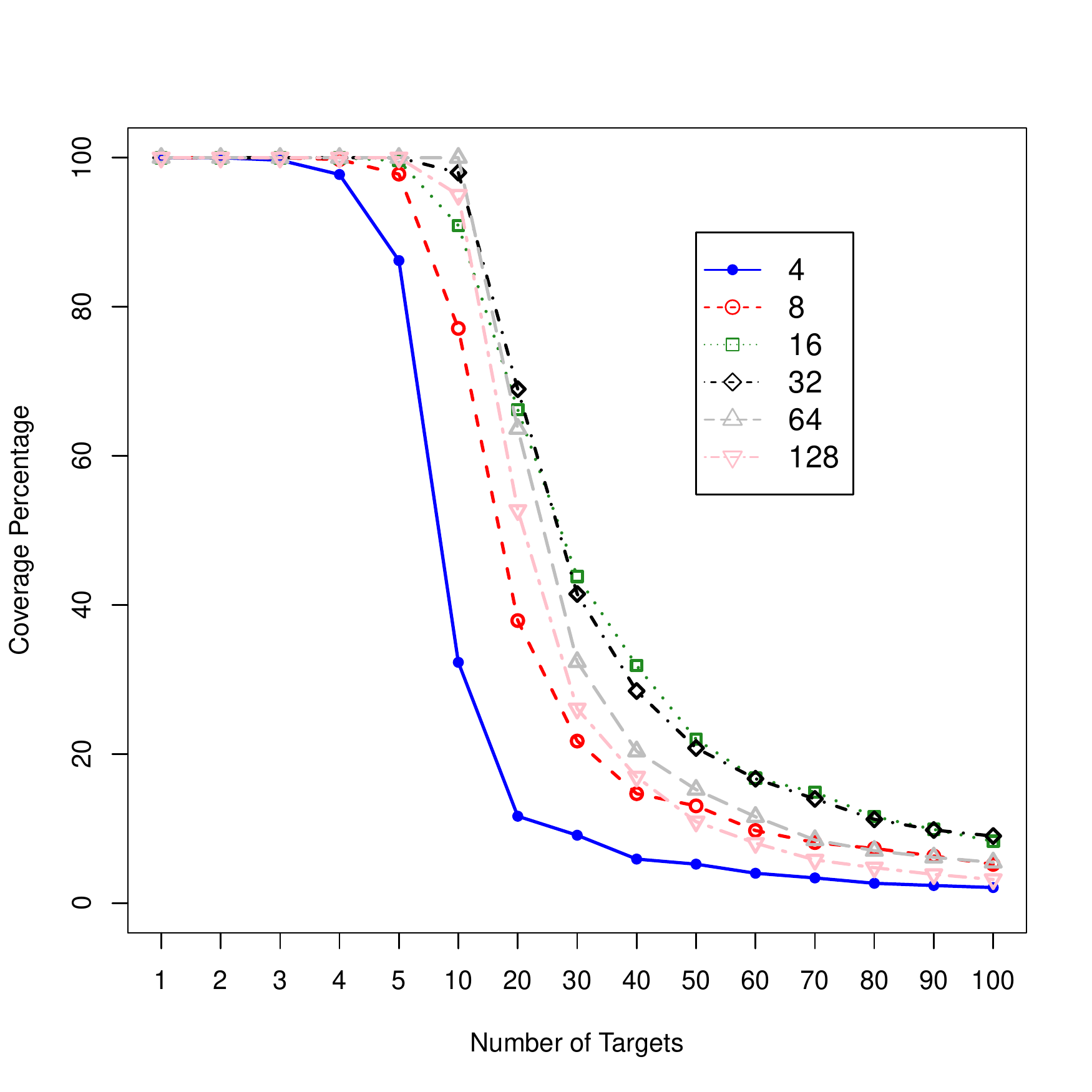}
  	\caption{\emph{Gradient} problem type.}
  \end{subfigure}\par\medskip
  \begin{subfigure}{\linewidth}
  	\includegraphics[width=.45\linewidth]{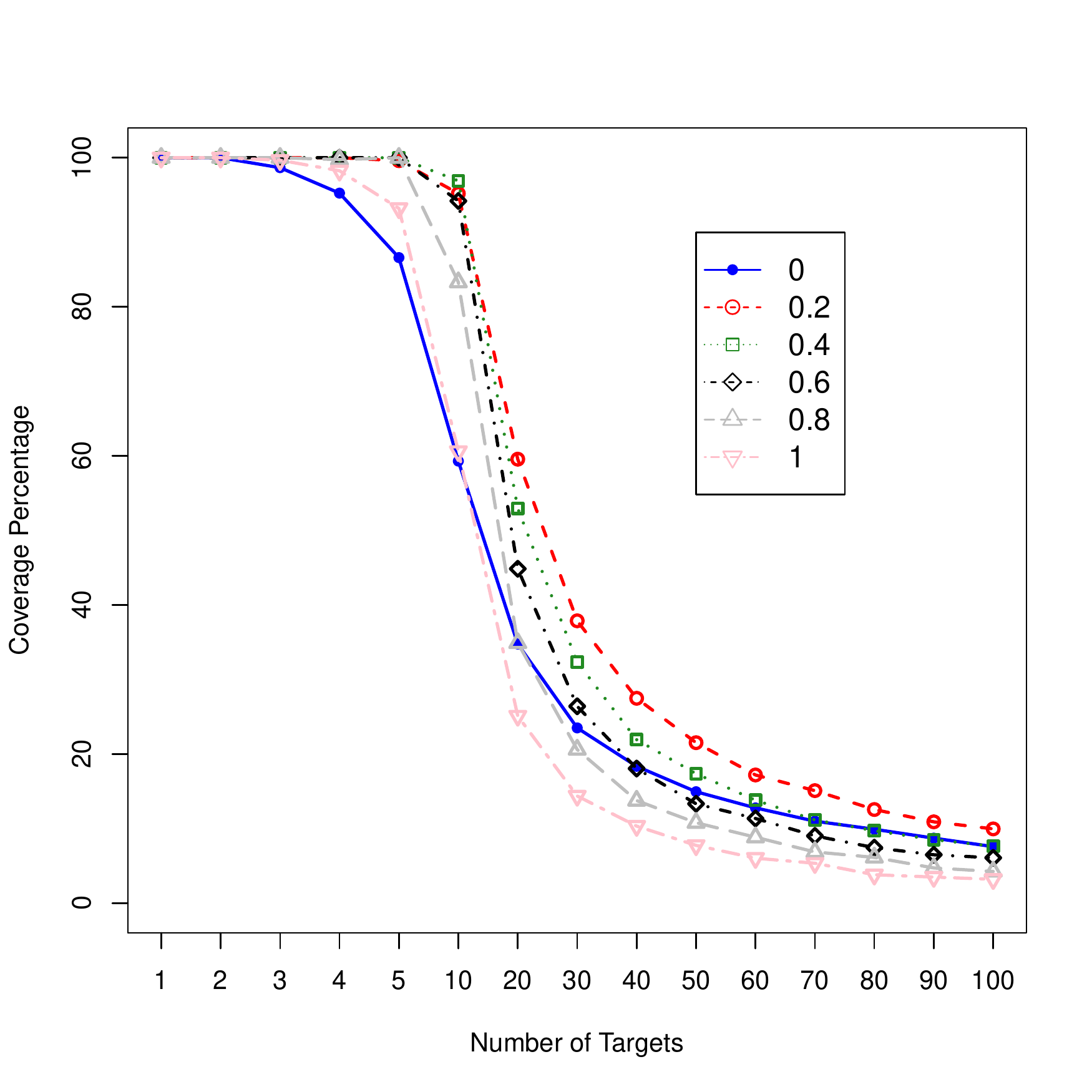}\hfill
  	\includegraphics[width=.45\linewidth]{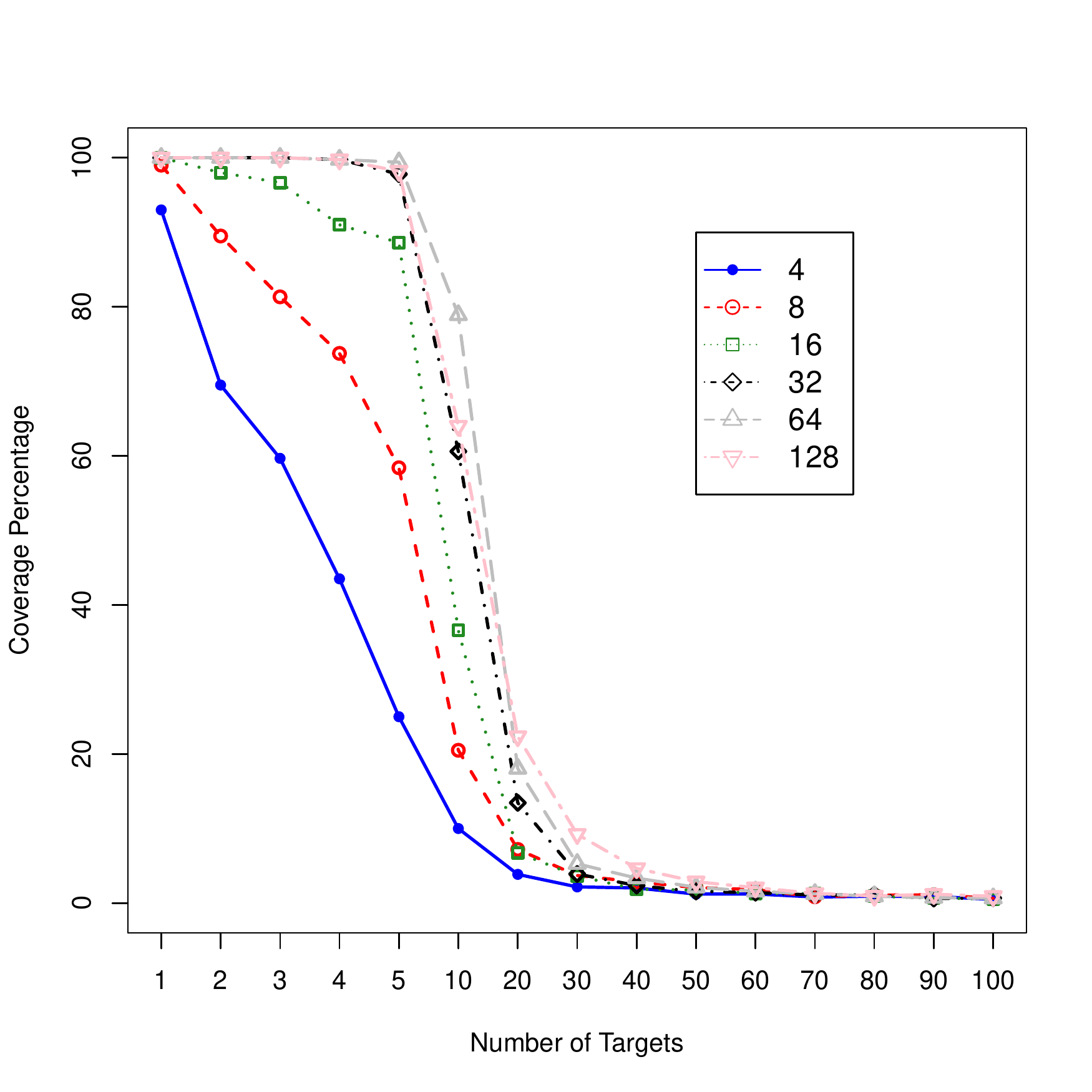}
  	\caption{\emph{Plateau} problem type.}
  \end{subfigure}\par\medskip
  \begin{subfigure}{\linewidth}
 	 \includegraphics[width=.45\linewidth]{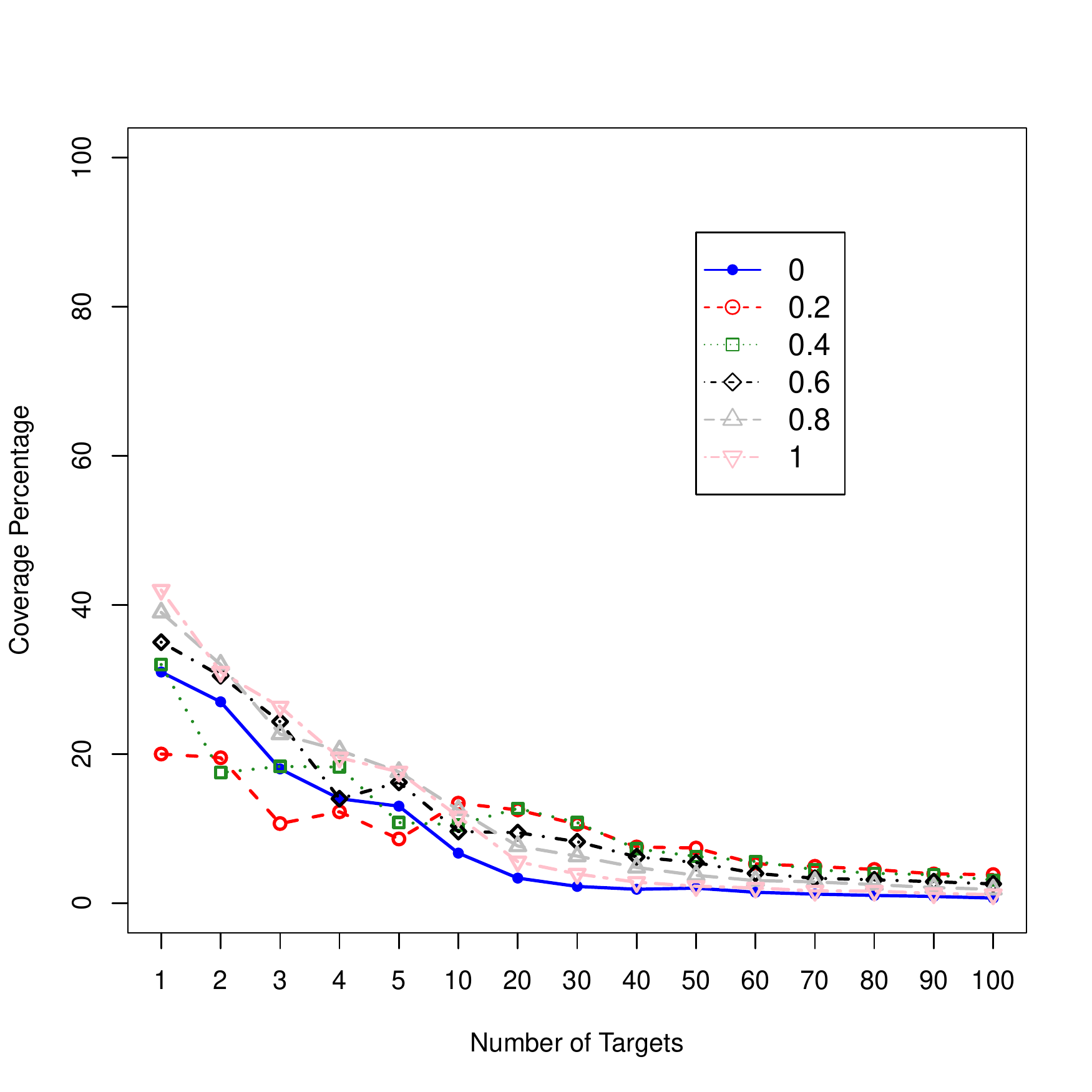}\hfill
  	\includegraphics[width=.45\linewidth]{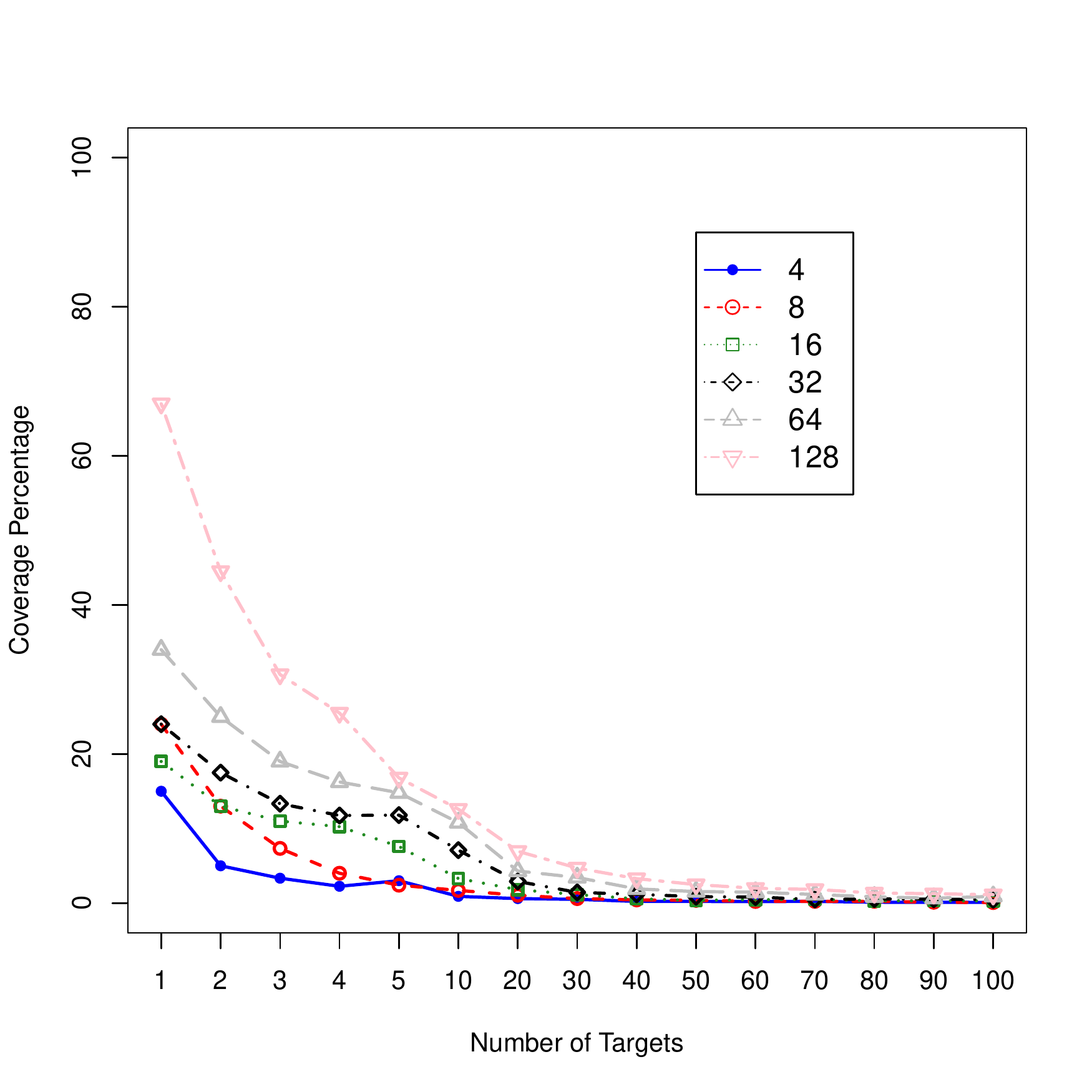}
 	 \caption{\emph{Deceptive} problem type.}
  \end{subfigure}
  \caption{\label{fig:tuning}
  Tuning of $F$ for MIO (left side) and population size for MOSA (right side).}
\end{figure}

To study these effects, we carried out a further series of experiments on the \emph{Gradient}, \emph{Plateau} and \emph{Deceptive} problem types.
For MIO, we studied six different values for $F$, in particular $\{0, 0.2, 0.4, 0.6, 0.8, 1\}$.
For MOSA, we studied six different values for the population size, i.e. $\{4, 8, 16, 32, 64, 128\}$.
Each experiment was repeated 100 times.
Figure~\ref{fig:tuning} shows the results of these experiments.

For MIO, the results in Figure~\ref{fig:tuning} do match expectation:
for problems with clear gradient or with just some plateaus, a more focused search that rewards exploitation is better.
The best setting is a low $F=0.2$, although the lowest $F=0$ is not particularly good. 
You still need some genetic diversity at the beginning of the search, and not rely on just one single individual.
For deceptive landscapes, exploration can be better, especially for a low number of targets.
For example, with $z=1$ then $F=1$ provides the best performance.
However, for larger number of targets, too much exploration would not be so beneficial, as it would not have enough time to converge to cover the targets.

In the case of MOSA, Figure~\ref{fig:tuning} provides some interesting insight.
For simple problems with clear gradient, one would expect that a focused search should provide better results.
However, the small population size of 4 is actually the configuration that gave the worst results.
The reason is that there is only little genetic material at the beginning of the search, and new one is only generated with the mutation operator.
However, a too large population size would still be detrimental, as not focused enough.
In that particular problem type, the best population size seems ranging from 16 to 32,
i.e., not too large, but not too small either.
In case of plateaus, still a too small population size (e.g., 4) gives the worst result.
However, in case of plateaus, there is a need to have some more exploration in the  search landscape, and this confirmed by the fact that the best results are obtained with large population sizes (e.g., 64 and 128).
This effect is much more marked in the case of deceptive landscapes, where large population sizes lead to much better results.

The experiments reported in Figure~\ref{fig:tuning} clearly points out to a challenge in population-based algorithms when dealing with many-objective problems.
A too small population size would reduce diversity in the initial genetic material.
But a too large population size would hamper convergence speed.
Finding a fixed, right population size that works on most problem sizes (e.g., $z=10$ vs $z=1m$) might not be feasible.
To overcome this issue, MIO uses a dynamically sized population, whereas the tradeoff between exploration and exploitation is controlled by a dynamically decreasing probability $P_r$ of creating new tests at random (instead of mutating the current ones stored in the archive).

\begin{result}
{\bf RQ2}: On the analysed problems, the population size and the $F$ parameter have clear effects on performance, which strongly depend on whether on the given problem one needs more or less exploitation/exploration.
\end{result}

%-----------------------------------------------------------------------
\subsection{Numerical Functions}

\begin{table}[!t]
\setlength{\tabcolsep}{3pt}
\centering
\caption{\label{tab:unit}
Comparions of algorithms on three different numerical functions.
Coverage is not a percentage, but rather the average raw sum of statements and branches that are covered.
For each algorithm, we also specify if better than any of the others, i.e. $\hat{A}_{12}>0.5$ (in parenthesis) and p-value less than $0.05$.
}
\begin{tabular}{ ll rr l} 
\toprule 
SUT & Algorithm & Tests & Coverage & Better than \\ 
\midrule 
\emph{Expint} & MIO & 9.4 & 63.7 &  RAND(1.00)  WTS(0.82) \\ 
 & MOSA & 14.0 & 63.2 &  RAND(1.00)  WTS(0.80) \\ 
 & RAND & 5.4 & 38.7 & \\ 
 & WTS & 9.3 & 62.5 &  RAND(1.00) \\ 
\midrule 
\emph{Gammq} & MIO & 9.2 & 69.1 &  MOSA(0.83)  RAND(1.00)  WTS(0.90) \\ 
 & MOSA & 8.0 & 65.9 &  RAND(1.00) \\ 
 & RAND & 1.0 & 32.0 & \\ 
 & WTS & 6.8 & 67.2 &  MOSA(0.60)  RAND(1.00) \\ 
\midrule 
\emph{Triangle} & MIO & 12.6 & 38.9 &  MOSA(0.71)  RAND(1.00)  WTS(0.98) \\ 
 & MOSA & 14.2 & 37.8 &  RAND(0.99)  WTS(0.86) \\ 
 & RAND & 11.1 & 31.7 & \\ 
 & WTS & 11.3 & 35.7 &  RAND(0.97) \\ 
\bottomrule 
\end{tabular} 

\end{table}

When designing algorithms to work on a large class of problems, it is common to evaluate them on artificial problems to  try to abstract away and analyse in details the characteristics for which such algorithms perform best.
For example, the very popular NSGA-II algorithm (on which MOSA is based on) was originally evaluated only on nine numerical functions~\cite{deb2002fast}.
However, using only artificial problems is risky, as those might abstract away some very important factors.
A good example of this issue is Adaptive Random Testing, where artificial problems with artificially high fault rates were masking away its very prohibitive computational cost~\cite{ArB11a}.

To somehow mitigate this issue, as a safety-net we also carried out some experiments on actual software, where we aim at unit testing for line and branch coverage.
We use the branch distance as heuristic for the fitness function.
We considered three numerical functions previously used in the literature (e.g., \cite{ArB11a}):
\emph{Expint} (88 LOC, including everything, also empty lines),
\emph{Gammq} (91 LOC),
and \emph{Triangle} (29 LOC).
Each algorithm was run for up to 5000 fitness evaluations.
Each experiment was repeated 100 times.
Average values are reported in Table~\ref{tab:unit},
where we also report the Vargha-Delaney effect sizes $\hat{A}_{12}$ and the results of 
Mann-Whitney-Wilcoxon U-tests at $\alpha=0.05$ level~\cite{Hitchhiker14}.
In all these three numerical functions, the MIO algorithm is the one achieving the highest coverage of the targets.
However, the three chosen numerical functions are not particularly difficult, and, as such, the performance difference between MIO, MOSA and WTS is not so large.

There is one thing to notice in these results:
WTS is much better than Random, whereas in the previous experiments they were very similar.
After an investigation, the reason for this behaviour is rather obvious.
With a population size of 50, and up to 50 tests in the same test suite, on average the first population would have a size of $50 \times 50/2 = 1250$ tests, which is higher than the search budget $b=1000$.
In other words, in those experiments WTS was practically doing just a random search.
However, this is not the case here, as we have $b=5000$.
In retrospective, on one hand those experiments could be considered unfair to WTS.
On the other hand, this issue further stresses out the need for a dynamically sized population when dealing with many-objective problems.

\begin{result}
{\bf RQ3}: the experiments on actual software are consistent with the ones on artificial problems: the MIO algorithm still achieves the best results.
\end{result}

%%%%%%%%%%%%%%%%%%%%%%%%%%%%%%%%%%%%%%%%%%%%%%%%%%%%%%%%%%%%%%%%%%%%%%%%%
% No space for it :(
%\section{Threats to Validity}
%\vspace{-0.5em}

%%%%%%%%%%%%%%%%%%%%%%%%%%%%%%%%%%%%%%%%%%%%%%%%%%%%%%%%%%%%%%%%%%%%%%%%%
\section{Conclusion}
%\vspace{-0.5em}

In this paper, we have presented a novel search algorithm that is tailored for the problem of generating test suites.
We call it the Many Independent Objective (MIO) algorithm.
We have carried out an empirical study to compare MIO with the other main algorithms for test suite generation: 
the Whole Test Suite (WTS) approach and the Many-Objective Sorting Algorithm (MOSA).
We also used random search as a baseline.
On artificial problems with increasing complexity and on some numerical functions, MIO achieved better results than the other algorithms.
In some cases, coverage improvements were even in the order of $+80\%$.

Future work will focus on implementing the MIO algorithm in different test generation frameworks, especially in system-level testing, and empirically evaluate how it fares in those contexts.
To help researchers integrate MIO in their frameworks, all the code used for the experiments in this paper is available online on a public repository, as part of the {\sc EvoMaster} tool at
\texttt{www.evomaster.org}

% ---------------------------------------------------- Acknowledgments
\vspace{0.5em}
\noindent {\bf Acknowledgments. }
This work is supported by the National Research Fund, Luxembourg (FNR/P10/03).

% ------------------------------------------------------- Bibliography

\bibliographystyle{splncs03}
%\bibliography{../../papers}

% that's all folks
\end{document}